\documentclass[preprint2]{aastex}

\newcommand{\vdag}{(v)^\dagger}

\slugcomment{Accepted for publication in the Astrophysical Journal 
}

\shorttitle{Electron Capture and $\beta$-decay Rates for O-Ne-Mg Core Stars}
\shortauthors{Suzuki, Toki, \& Nomoto}

\begin{document}


\title{Electron capture and $\beta$-decay rates for $sd$-shell nuclei in stellar environments relevant to high density O-Ne-Mg cores}


\author{Toshio Suzuki\altaffilmark{1}}

\affil{Department of Physics and Graduate School of Integrated
Basic Sciences,\\ 
College of Humanities and
Sciences, Nihon University\\
Sakurajosui 3-25-40, Setagaya-ku, Tokyo 156-8550, Japan}
\email{suzuki@phys.chs.nihon-u.ac.jp}

\author{Hiroshi Toki}
\affil{Research Center for Nuclear Physics (RCNP), Osaka University,\\ Ibaraki,
Osaka 567-0047, Japan}

\and

\author{Ken'ichi Nomoto\altaffilmark{2}}
\affil{Kavli Institute for the Physics and Mathematics of the Universe (WPI),\\
The University of Tokyo, Kashiwa, Chiba 277-8583, Japan\\
\vspace*{1.0cm}
{\rm Accepted for publication in the Astrophysical Journal}}

\altaffiltext{1}{Visiting Researcher, National Astronomical Observatory of Japan, Mitaka, Tolyo 181-8588, Japan}
\altaffiltext{2}{Hamamatsu Professor}



\begin{abstract}
Electron capture and $\beta$-decay rates for nuclear pairs in $sd$-shell are evaluated at
high densities and high temperatures relevant to the final evolution of electron-degenerate
O-Ne-Mg cores of stars with the initial masses of 8-10 M$_{\odot}$. 
Electron capture induces a rapid contraction of the
electron-degenerate O-Ne-Mg core.
The outcome of rapid contraction depends
on the evolutionary changes in the central density and temperature,
which are determined by the competing processes of contraction,
cooling, and heating.  
The fate of the stars are determined by these competitions, 
whether they end up with electron-capture
supernovae or Fe core-collapse supernovae. 
Since the competing processes are induced by electron capture and $\beta$-decay, 
the accurate weak rates are crucially important. 
The rates are obtained for pairs with A=20, 23, 24, 25 and 27 by shell-model calculations
in $sd$-shell with the USDB Hamiltonian.
Effects of Coulomb corrections on the rates are evaluated.
The rates for pairs with A=23 and 25 are important for nuclear URCA processes
that determine the
cooling rate of O-Ne-Mg core, while those
for pairs with A=20 and 24 are important for the core-contraction 
and heat generation rates in the core.
We provide these nuclear rates
at stellar environments in tables with fine enough meshes at various densities and
temperatures for the studies of astrophysical
processes sensitive to the rates. 
In particular, the accurate rate tables are crucially important 
for the final fates of
not only O-Ne-Mg cores but also a wider
range of stars such as C-O cores of lower mass stars. 
\end{abstract}





\section{Introduction}

The evolution and final fates of stars depend on their initial masses $M_{\rm I}$ 
\citep[e.g.,][]{Nomoto13}, 
which is still subject to some uncertainties involved in stellar mass-loss, mixing processes
and nuclear transition rates.
Stars with $M_{\rm I} =$ 0.5--8\,M$_{\odot}$ form a C-O core after He burning and
become C+O white dwarfs.
Stars with $M_{\rm I} >$ 10\,M$_{\odot}$, on the other hand,
form an Fe core and later explode as core-collapse supernovae.
In stars with $M_{\rm I} =$ 8--10\,M$_{\odot}$, an O-Ne-Mg core is formed after carbon
burning. 
8--10\,M$_\odot$ stars can end up in various ways, that is,
as O-Ne-Mg white dwarfs without
explosions, or as electron-capture supernovae or Fe core-collapse supernovae
\citep{Nomoto, Miyaji, Nomoto2, Nomoto3}.

The evolutionary changes in the central density and temperature of the O-Ne-Mg core are determined
by the competition among the contraction, cooling and heating processes.
If heating is fast enough relative to
contraction, the temperature would become high enough to ignite
Ne-O burning, which would result in the Fe core formation.
If cooling is fast, the contraction would lead to the collapse of the
O-Ne-Mg core and an electron-capture supernova.
The fate of the stars
with $M_{\rm I} \sim$ 8-10 M$_{\odot}$ is thus determined by these competitions.
Since these competing processes are induced by electron capture and $\beta$-decay, the accurate
weak rates are crucially important.

We study electron capture and $\beta$-decay rates in $sd$-shell nuclei at stellar
environments relevant to the final evolution of electron-degenerate O-Ne-Mg cores
in stars with the initial masses of $M_{\rm I} =$ 8--10\,M$_{\odot}$.
Nuclear URCA processes, especially in nuclear pairs with A=23 and 25, are important
for the cooling of the O-Ne-Mg cores after carbon burning \citep{TSN,Jones}
Electron capture reactions and successive gamma emissions in nuclei with A=24 and 20
are important for the contraction and heating of the core in later stages leading to 
an electron-capture supernova. 
The cooling and heating rates of the core as well as the core-contraction rate are 
determined by the weak rates for the nuclear pairs. 
The fate of the stars depend sensitively on the nuclear electron capture and $\beta$-decay rates.

Precise evaluations of the weak rates in stars are quite important for
the evolutions of not only the stars having O-Ne-Mg cores but also lower mass stars having
C-O cores (which contain primordial species of A = 20-27). 
The rates are also important for the final fates of O-Ne-Mg, C-O and hybrid C-O-Ne white
dwarfs in binary systems. 
We thus provide electron capture and $\beta$-decay rates at stellar environments with such fine
meshes as $\Delta$ log$_{10}(\rho Y_e)$ =0.02 
($Y_e$ is the electron mole number and $\rho$ is the nucleon density in units of g cm$^{-3}$)
and  $\Delta$ log$_{10}T$ =0.05 ($T$ is the temperature in units of K) 
at various densities and temperatures, 
so that they can be used for the studies of astrophysical processes 
sensitive to the rates as well as the determination of the final fates of stars.
We provide the Table for the weak rates on the web \citep{Tab}.      

In sect. 2, Q-values in $sd$-shell nuclei and relations to triggering electron capture
reactions are discussed.
In sect. 3, formalism for electron capture and $\beta$-decay rates as well as neutrino energy-loss
rate and $\gamma$-ray heating rate are given.
Coulomb corrections are also taken into account.
In sect. 4, the weak rates and energy loss by neutrino emissions for nuclear pairs
with A=23, 25 and 27,
which are important for the URCA processes, are studied.
Summary is given in sect. 5.

\section{Q-values in $sd$-shell nuclei}

The triggering of electron capture reactions is determined by Q-values of nuclear transitions
and values of the chemical potential of electrons at high densities and high temperatures.
The Q-values for $\beta$-decays in $sd$-shell nuclei with A=17-31 are shown in Fig. 1.
They are taken from \citet{Ames2012}.

\begin{figure*}[tbh]
\includegraphics[scale=0.92]{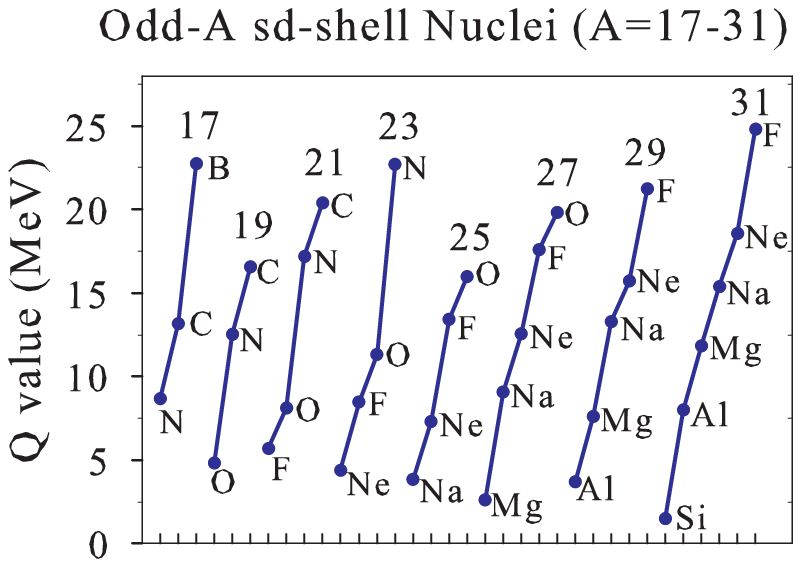}
\includegraphics[scale=0.95]{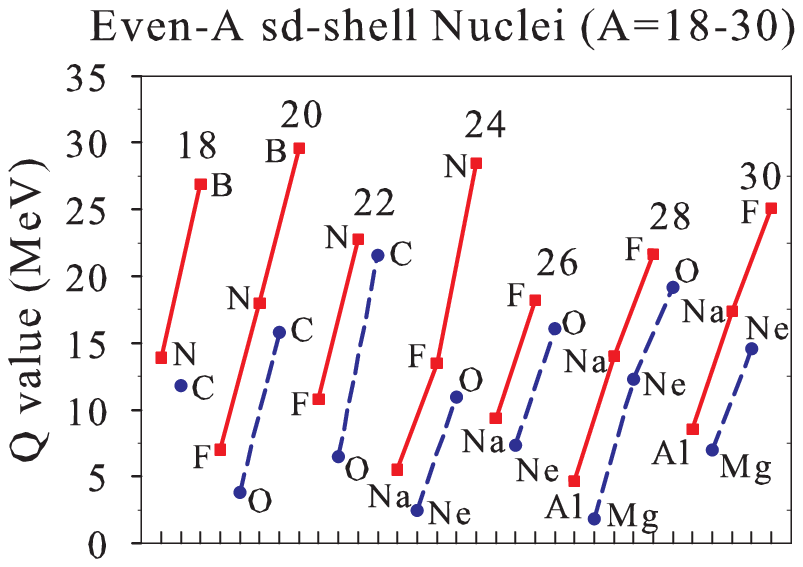}
\caption{
Q-values for $\beta$-decays for $sd$-shell nuclei with odd-A (A=17-31) and even-A
(A=18-30). Parent nuclei with proton number Z and mass number A for $\beta$-decays, (Z, A) $\rightarrow$ (Z+1, A) + e$^{-}$ +$\bar{\nu}_e$, 
are denoted in the figures. Data are taken from \citet{Ames2012}. For even-A case, transitions
from odd-odd and even-even nuclei are shown by solid and dashed lines, respectively.
\label{fig:fig1}}
\end{figure*}

In general, the Q-values can become small for odd-A nulei compared to even-A nuclei as
can be noticed from Fig. 1. They are particularly as small as 1.5 -4.4 MeV for the
pairs with A =23, 25, 27, 29 and 31; $^{23}$Ne $\rightarrow$ $^{23}$Na, $^{25}$Na
$\rightarrow$ $^{25}$Mg, $^{27}$Mg $\rightarrow$ $^{27}$Al, $^{29}$Al $\rightarrow$
$^{29}$Si and $^{31}$Si $\rightarrow$
$^{31}$P.
The URCA processes can occur for these nuclear pairs if the transition from the ground
state (g.s.) of the mother nucleus to the ground state of the daughter nucleus is not
forbidden.
In fact, the URCA processes for the pairs, $^{23}$Ne $\leftrightarrow$ $^{23}$Na, $^{25}$Na
$\leftrightarrow$ $^{25}$Mg, are found to be important for the cooling of the
O-Ne-Mg core of stars with 8--10\,M$_{\odot}$ \citep{TSN,Jones}.
As the g.s. to g.s. transition for $^{27}$Mg $\leftrightarrow$ $^{27}$Al is a forbidden
transition, the URCA process does not occur for this pair \citep{TSN} though the Q-value
is very small; Q=2.61 MeV.
The order of the occurrence of the URCA processes depends on the Q-value of the transition.
The URCA processes for the pair $^{25}$Na-$^{25}$Mg with Q=3.83 MeV takes place first,
and it is followed by the URCA process for the pair $^{23}$Ne-$^{23}$Na with Q=4.38 MeV.

For even-A nuclei, the $\beta$-decay Q-values are generally large for transitions from odd-odd
nuclei to even-even nuclei while they are relatively small for transitions from even-even
to odd-odd nuclei owing to the pairing effects.
In later stages of the evolution of O-Ne-Mg core stars, electron capture reactions on even-even nuclei
such as $^{24}$Mg with
Q-value of 5.52 MeV and
$^{20}$Ne with Q-value of 7.02 MeV are triggered at higher densities.
They are succeeded by electron capture reactions on $^{24}$Na and $^{20}$F, respectively.
These odd-odd nuclei have small threshold energies for the electron capture reactions, 2.47 MeV for
$^{24}$Na and 3.81 MeV for $^{20}$F.
The double electron capture reactions on $^{24}$Mg and $^{20}$Ne become
important for heating stars in later stages of star evolutions. 

Here, we consider cooling of O-Ne-Mg core by URCA processes in nuclear pairs with A =23, 25 and 27,
and heating of stars by electron capture reactions on nuclei with A =20 and 24.
The cooling process occurs at log$_{10}(\rho Y_e)$ =8.8-9.0 first for $^{25}$Na -$^{25}$Mg and
$^{23}$Ne -$^{23}$Na pairs, and then secondary one occurs at log$_{10}(\rho Y_e)$ =9.3-9.8 for
other pairs with the second smallest Q-values in each mass number A =23, 25 and 27 \citep{Sam13}.
The heating process occurs at log$_{10}(\rho Y_e)$ =9.2-9.7 for captures on $^{24}$Mg and $^{20}$Ne.
The successive electron captures on $^{24}$Na and $^{20}$F with smaller Q-values occur at
log$_{10}(\rho Y_e)$ =8.2 and 8.8.
In the present work, the electron capture and $\beta$-decay rates in these $sd$-shell nuclei which concern
the cooling of O-Ne-Mg core after C burning and the heating before O burning are evaluated
covering the density region mentioned above, log$_{10}(\rho Y_e)$ = 8.0 -10.0, for relevant temperatures.
The weak rates for nuclei with A =17-28 as well as neutrino energy-loss rates and $\gamma$-ray heating rates
are obtained for densities log$_{10}(\rho Y_e)$ = 8.0 -11.0 in fine steps of 0.02, and temperatures
log$_{10}T$ = 8.0 -9.65 and 7.0 -8.0 in steps of 0.05 and 0.20, respectively \citep{Tab}.

\section{Formulation for the weak reaction rates and Coulomb effects}

The electron capture and $\beta$-decay transitions are dominantly induced by GT
transitions in the present stellar conditions with densities log$_{10}(\rho Y_e)$ = 8 -10
and temperatures log$_{10}T$ = 8 -9.6.
Structure of $sd$-shell nuclei are well described by a shell-model Hamiltonian,
USD \citep{BW83,BW87,BW88}. Energy spectra, magnetic dipole (M1) and Gamow-Teller (GT)
strengths with quenched spin g-factor and axial-vector coupling constant, $g_A$,
as well as Q moments and E2 transition strengths with effective charges are
systematically well reproduced with USD. The Hamiltonian has been updated by
taking into account recent data. While neutron-rich O and F isotopes were too tightly
bound for USD, the new versions, USDA and USDB \citep{BR,RB}, are free from this
problem. Here, USDB is used for the evaluation of $B(GT)$ with $g_A^{\rm eff}$/$g_A$ = 0.764.

Electron capture rates at high densities and high temperatures are evaluated as 
\citep{Full80,Full82a,Full82b,Full85,Lan1,Lan2,SHM}

\begin{eqnarray}
\lambda &=& \frac{\rm ln2}{6146 (s)} \sum_{i} W_i \sum_{f}
(B_{if}(GT) \nonumber\\ 
&+& B_{if}(F)) \Phi^{ec}(Q_{if})) 
\nonumber\\
\Phi^{\rm ec}(Q_{if}) &=& \int_{\omega_{min}}^{\infty} \omega p (Q_{if}+\omega)^2
F(Z, \omega) S_e(\omega) d\omega, \nonumber\\
Q_{if} &=& (M_{\rm p}c^2 -M_{\rm d}c^2 +E_i -E_f)/m_{e}c^2, \nonumber\\
W_i &=& (2J_i +1) e^{-E_i/kT} \nonumber\\
& &/\sum_{i}(2J_i +1) e^{-E_i/kT},
\end{eqnarray}
where $\omega$ and $p$ are electron energy and momentum
in units of
$m_{e}c^2$ and $m_{e}c$, $M_{\rm p}$ and $M_{\rm d}$ are nuclear mass of parent and daughter nuclei,
respectively, and $E_i$, $E_f$ are excitation energies of initial and final states.
$J_i$ is the total spin of initial state.
Here, $B(GT)$ and $B(F)$ are the GT and Fermi strengths, respectively, given by
\begin{eqnarray}
B_{if}(GT) &=& (g_{A}/g_{V})^2 \frac{1}{2 J_i+1}  |\langle f||\sum_k \sigma^k t_{+}^{k} ||i\rangle|^2 \nonumber\\
B_{if}(F) &=& \frac{1}{2 J_i+1}  |\langle f||\sum_k t_{+}^{k} ||i\rangle|^2
\end{eqnarray}
where 
$t_{+}|p>=|n>$.
In case of $\beta$-decay, $t_{+}$ is replaced by $t_{-}$; $t_{-}|n>=|p>$.
$F(Z, \omega)$ is the Fermi function, and $S_e(\omega)$ is the Fermi-Dirac
distribution for electrons where the chemical potential, $\mu_{e}$,
is determined from the density, $\rho Y_e$,
by
\begin{equation}
\rho Y_e = \frac{1}{\pi^{2}N_{A}}(\frac{m_{e}c}{\hbar})^3
\int_{0}^{\infty}(S_{e} -S_{p})p^2dp
\end{equation}
where $N_A$ is the Avogadro number and $S_{p}$ is the Fermi-Dirac distribution
for positrons with the chemical potential $\mu_{p}$ = $-\mu_{e}$.
It can become as large as 2 -5 (11) MeV at high densities log$_{10}(\rho Y_e)$ = 8 -9 (10).
It slightly decreases as the temperature increases.
The reaction rates become larger at higher densities because of the large
chemical potential.

Since it is sometimes important to include GT transitions from thermally
populated excited states in steller interior, we include
transitions from the excited states of the parent nucleus in addition to
the ground state with the partition function $W_i$.
When temperature becomes high and excitation energies of excited states
of the parent nucleus are low, the nucleus can be thermally excited and
the population of the excited states can be large. In such a case, the
transitions from the excited states can give important contributions to
the capture rates.
In the present work, excited states with excitation energies up to $E_x$ = 2 MeV
are taken into account.

In case of $\beta$-decays, the integral in Eq. (1) is replaced by
\begin{equation}
\Phi^{\beta}(Q_{if}) =\int_{1}^{Q_{if}} \omega p (Q_{if}-\omega)^2
F(Z+1, \omega) (1-S_e(\omega)) d\omega
\end{equation}
The $\beta$-decay rates decrease as the density increases due to the blocking of the
electron density in contrary to the electron capture case.
There is, therefore, a density where both the electron capture and $\beta$-decay rates are
balanced. 
When such densities depend little on tempertures, both the transitions
occur simultaneously emitting both $\nu$ and $\bar{\nu}$
at such densities, which we call "URCA densities".
Here, the rates are evaluated for 8.0 $<$ log$_{10}$($\rho Y_e$) $<$11.0 in fine steps of 0.02
and 8.0 $<$ log$_{10}T$ $<$ 9.65 in steps of 0.05 and also for 7.0 $<$ log$_{10}T$ $<$8.0 in steps of 0.20.

We give also formulae for the energy loss  by neutrino emissions and heating by $\gamma$
emissions for astrophysical applications.
The neutrino energy-loss rate is given by
\begin{eqnarray}
\xi &=& \frac{(\rm ln2) m_{e}c^2}{6146 (s)} \sum_{i} W_i \sum_{f}
(B_{if}(GT) \nonumber\\
&+& B_{if}(F)) \Psi^{ec}(Q_{if}))
\nonumber\\
\Psi^{\rm ec}(Q_{if}) &=& \int_{\omega_{min}}^{\infty} \omega p (Q_{if}+\omega)^3
F(Z, \omega) \nonumber\\
& &\times S_e(\omega) d\omega
\end{eqnarray}
for electron capture reactions, while for $\beta$-decay transitions $\Psi^{ec}$ is replaced by
\begin{equation}
\Psi^{\beta}(Q_{if}) =\int_{1}^{Q_{if}} \omega p (Q_{if}-\omega)^3
F(Z+1, \omega) (1-S_e(\omega)) d\omega.
\end{equation}
The average energy of the emitted neutrino is obtained by the ratio
\begin{equation}
\langle E_{\nu} \rangle = \frac{\xi}{\lambda}.
\end{equation}

The $\gamma$-ray heating rate is given by \citep{Oda,Taka}
\begin{eqnarray}
\eta &=& \frac{(\rm ln2) m_{e}c^2}{6146 (s)} \sum_{i} W_i \sum_{f}
(B_{if}(GT) \nonumber\\
&+& B_{if}(F)) \Gamma^{ec}(Q_{if}))
\nonumber\\
\Gamma^{\rm ec}(Q_{if}) &=& \int_{\omega_{min}}^{\infty} \omega p (Q_{if}+\omega)^2
E_{f}/m_{e}c^2 F(Z, \omega) \nonumber\\
& &\times S_e(\omega) d\omega
\end{eqnarray}
for electron capture reactions, while for $\beta$-decay transitions $\Gamma^{ec}$ is replaced by
\begin{eqnarray}
\Gamma^{\beta}(Q_{if}) &=& \int_{1}^{Q_{if}} \omega p (Q_{if}-\omega)^2
E_{f}/m_{e}c^2 \nonumber\\
& &\times F(Z+1, \omega)(1-S_e(\omega)) d\omega.
\end{eqnarray}
The average energy of the emitted $\gamma$ is obtained by the ratio
\begin{equation}
\langle E_{\gamma} \rangle = \frac{\eta}{\lambda}.
\end{equation}
Here, the excited states of the daughter nucleus are assumed to emit $\gamma$'s leading
to its ground state \citep{Oda,Taka}.

Next, the Coulomb corrections on the transition rates due to the electron background
are studied.
There are two Coulomb effects; one is the screening effects of electrons and the other
is the change of threshold energy due to the change of chemical potential of
the ions.

The Coulomb potential is modified due to the screening effects of relativistically
degenerate electron liquid. The modification
can be evaluated by using the dielectric function obtained by random phase
approximation \citep{Itoh}. This effect is included by reducing the chemical potential
of electrons by an amount equal to the modification of the Coulomb potential at the origin,
$V_s(0)$ \citep{Juod}, where
\begin{eqnarray}
V_{s}(r)
&=& Ze^{2}(2k_{F})J(r) \nonumber\\
J(r) &=& \frac{1}{2 k_{F}r} \left(1- \frac{2}{\pi}\int \frac{\sin(2k_{F}qr)}{q^{2}\epsilon(q,0)}dq\right).
\end{eqnarray}
The screening coefficient $J$ is tabulated in \citet{Itoh}
(see \citet{TSN} for more details).
The electron capture and $\beta$-decay rates are slightly reduced
and enhanced, respectively, by the effect.

The other effect is caused by the correction of the chemical potential of the nucleus
with charge number Z, $\mu_{\rm C}(Z)$, due to the interactions of the nucleus with the electron
background \citep{Slat,Ichi}.
The threshold energy changes by
\begin{equation}
\Delta Q_{\rm C} = \mu_{\rm C}(Z-1) -\mu_{\rm C}(Z)
\end{equation}
which gets larger for electron capture processes.
The Coulomb chemical potential of a nucleus with $Z$ in a plasma of
electron number density $n_e$ and temperature $T$ is given by
\begin{equation}
\mu_{\rm C}(Z) = kT f(\Gamma)
\end{equation}
with $\Gamma$ = $Z^{5/3}\Gamma_e$, $\Gamma_e$ = $\frac{e^2}{kTa_e}$ and $a_e$ = $(\frac{3}{4\pi n_e})^{1/3}$.
The function $f$ for $\Gamma >$1 is given in \citet{Ichi}. 
More details are given in  \citet{TSN}.
This effect results in the reduction of the
electron capture rates and the enhancement of the $\beta$-decay rates.

Contribution of the second effect is larger than the first one, and thus the
electron capture ($\beta$-decay) rates are reduced (enhanced). The URCA density is
shifted to a larger density region as shown in the next section.

\section{Nuclear weak rates for pairs with A=23, 25 and 27}
\subsection{$^{23}$Ne -$^{23}$Na and $^{25}$Na -$^{25}$Mg pairs}

The electron capture and $\beta$-decay rates for A=23, 25 and 27 are evaluated at densities
and temperatures of stellar environments.
The rates for the pair, $^{25}$Na -$^{25}$Mg,  are shown in Fig. 2.
Transitions from 5/2$^{+}_{\rm g.s.}$, 1/2$^{+}$, 3/2$^{+}$, 7/2$^{+}$ and second 5/2$^{+}$ states
of $^{25}$Mg are included for the electron capture reactions, while 5/2$^{+}_{\rm g.s.}$, 3/2$^{+}$ and
1/2$^{+}$ states of $^{25}$Na are taken into account for the $\beta$-decays.
Here, the Coulomb effects are not included.
The electron capture rates increase as the density and
the chemical potential increases. The $\beta$-decay rates, on the other hand, decrease
as the density increases due to the blocking of the decays by the electrons with high
chemical potential.
The URCA density, where both the rates coinside, is found at log$_{10}(\rho Y_e)$ =8.77.
Dependence of the density on the temperature is quite small; 
$\Delta$log$_{10}{\rho Y_e} <$ 0.01 for log$_{10}T$ =8-9.2. 
The URCA density is also found at log$_{10}(\rho Y_e)$ =8.92 for the pair with A=23, $^{23}$Ne -$^{23}$Na. The temperature dependence is also as small as
$\Delta$log$_{10}{\rho Y_e} \approx$ 0.01 for log$_{10}T$ =8-9.2.
It is important to evaluate the rates with fine meshes to obtain the clear URCA densities.
Detailed discussion is given in \citet{TSN}.
In case of the pair with A=27, $^{27}$Mg -$^{27}$Al, where the GT transition from the
initial ground state to the final
ground state 
does not occur because of spin difference larger than 1, 
it is difficult to assign an URCA density 
due to the Q-value mismatch in the two transitions.
The densities where e-capture and $\beta$-decay rates coincide depend much
on the temperature for the A=27 case.

\begin{figure*}[tbh]
\includegraphics[scale=1.50]{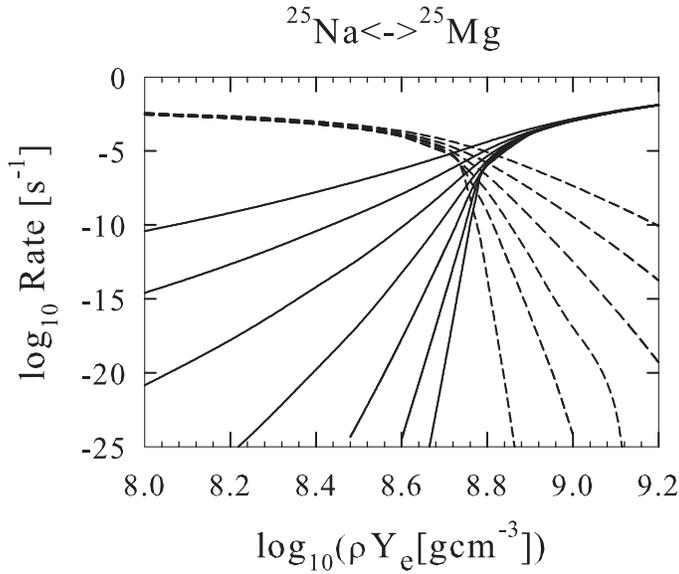}
\caption{
$\beta$-decay and electron capture rates for the A=25 URCA nuclear pairs,
$^{25}$Na -$^{25}$Mg, for various temperatures, log$_{10}T$ = 8-9.2 in steps
of 0.2, as functions of density
log$_{10}(\rho Y_e)$.
$\beta$-decay rates (dashed curves) decrease with density
while electron capture
rates (solid curves) increase with density.
\label{fig:fig2}}
\end{figure*}

Here, we comment on the choice of the steps of the mesh in 
log$_{10}(\rho Y_e)$ and log$_{10}T$. 
Fine steps of 0.02 and 0.05 are adopted for both log$_{10}(\rho Y_e)$ and
log$_{10}T$, respectively, as it is not easy to obtain accurate rate values 
from the interpolation procedure among quantities which
differ by many orders of magnitude. 
For example, from the rates at log$_{10}(\rho Y_e)$ = 7, 8, 9, 10 and 11 only
for a fixed temperature, it is not possible to get an accurate rate by
interpolation procedures. The error in the rate can be larger than a factor
of 10.
 
It is true that a procedure using effective log {\it ft} values proposed by
\citet{Full85} works well for certain cases. The procedure is based on the
fact that the change of the rates by orders of magnitudes comes mainly from
the phase space factor, and the remaining part including the nuclear transition 
probability does not change drastically. 
As the phase space factor for the transition between ground states are taken in the procedure, the effective log {\it ft} method is valid for the cases 
where GT transitions between the ground states are dominant. 
In case that the g.s. to g.s. transitions are forbidden or transitions from
excited states give essential contributions, the method becomes invalid 
when it is applied to the interpolation with sparce grid of densities at
log$_{10}(\rho Y_e)$ =7, 8, 9, 10 and 11.
The pairs $^{27}$Al -$^{27}$Mg and $^{20}$F -$^{20}$Ne are such examples.   

In a density interval of $\Delta$log$_{10}(\rho Y_e)$ =0.02 or 
in a temperature interval of $\Delta$log$_{10}T$ =0.05, the rate 
(log$_{10}$(rate)) can change by 3-4 orders of magnitude (by 3-4)
at most (see Fig. 2). For this order of the change in log$_{10}$(rate)
in the relevant intervals,
the interpolation can be safely done with an accuracy of
$\Delta$log$_{10}$(rate) =0.001-0.002.
The steps of the meshes chosen here are sufficient for obtaining the rates
with further finer meshes by the interpolation procedures.

\begin{figure*}[bht]
\hspace*{8mm}
\includegraphics[scale=1.3]{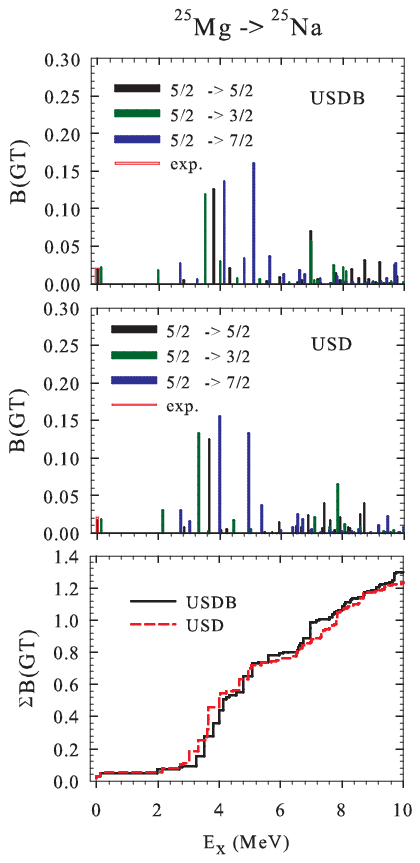}
\hspace*{8mm}
\includegraphics[scale=1.3]{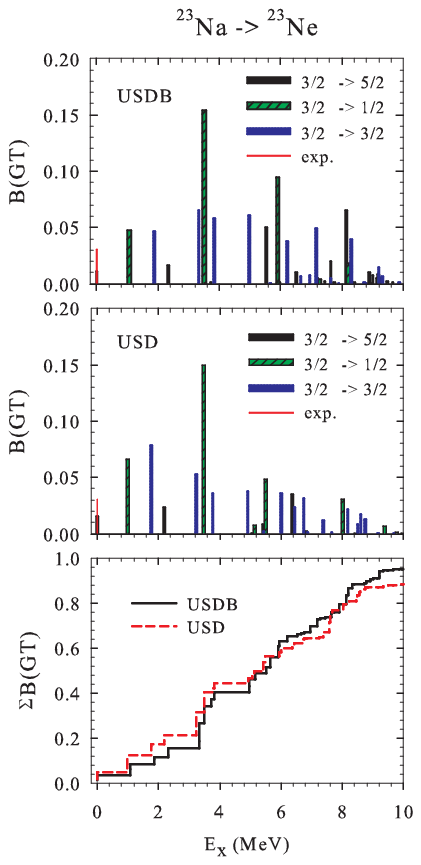}
\caption{
GT strengths and their cumulative sums for $^{25}$Mg and $^{23}$Na 
obtained with USDB and USD as well as the experimental values for 
the transition to the ground states of the daughter nuclei.
\label{fig:fig3}}
\end{figure*}

\begin{deluxetable}{ccccccccccc}
\tablecolumns{11}
\tablecaption{Transition Rates for $^{23}$Na-$^{23}$Ne Pair.
USDB* denotes that available experimental data are taken into account.
Values for USD* are taken from \citet{Oda}. }
\tablehead{
\colhead{}  & \colhead{} & \multicolumn{4}{c}{electron-capture rates} &
\colhead{} &
\multicolumn{4}{c}{$\beta$-decay rates} \\
\cline{3-6} \cline{8-11} \\
\colhead{$\rho Y_e$} & \colhead{$T/10^{9}$} & \colhead{USDB} &
\colhead{USDB*}  & \colhead{USD} & \colhead{USD*} &
\colhead{}  & \colhead{USDB} & \colhead{USDB*} & \colhead{USD} &
\colhead{USD*}}
\startdata
$10^{8}$ & 0.1 & \nodata & \nodata & \nodata & \nodata &
& -2.418 & -2.048 & -2.284 & -2.064\\
 & 0.4 & -37.243 & -36.899 & -37.108 & -36.913 &
& -2.417 & -2.048 & -2.284 & -2.063 \\
 & 1.0 & -17.640 & -17.261 & -17.505 & -17.278 &
& -2.414 & -2.044 & -2.280 & -2.060 \\
 & 2.0 & -10.644 & -10.256 & -10.508 & -10.270 &
& -2.395 & -2.028 & -2.257 & -2.042 \\
 $10^{9}$ & 0.1 & -4.762 & -4.333 & -4.626 & -4.352 &
& -24.294 & -23.866 & -24.158 & -23.185 \\
 & 0.4 & -4.710 & -4.282 & -4.574 & -4.299 &
& -10.700 & -10.271 & -10.564 & -10.290 \\
 & 1.0 & -4.461 & -4.046 & -4.324 & -4.064 &
& -7.265 & -6.854 & -7.102 & -6.863 \\
 & 2.0 & -3.965 & -3.580 & -3.824 & -3.597 &
& -5.275 & -5.015 & -5.073 & -4.983 \\
\enddata
\end{deluxetable}

Next, we compare the present results obtained with USDB with those of
\citet{Oda}. In \citet{Oda}, USD interaction was used and experimental
energies and GT transition rates available were also taken into account
\citep{BW85}.  
Calculated GT strengths as well as their cumulative sums for USDB and USD
are shown in Fig. 3 for the transitions $^{25}$Mg $\rightarrow$ $^{25}$Na
and $^{23}$Na $\rightarrow$ $^{23}$Ne.
Experimental transition strengths to the ground states of the daughter nuclei
are also shown.
In both $^{25}$Mg and $^{23}$Na, the GT strength is more spread for USDB
compared with USD and larger strength remains more in the higher excitation
energy ($E_x$) region for USDB.  
The GT strengths differ little at low excitation energy region $E_x <$ 3 MeV 
for $^{25}$Mg, while for $^{23}$Na
their difference in the magnitude is noticed at low $E_x$ region below 3 MeV. 
Calculated $B(GT)$ values for the transition to the  ground state of $^{25}$Na
for both USDB and USD are consistent with the experimental value.
The latter is only a bit larger than the values of USDB (USD) by
15$\%$ (19$\%$).
Transition  rates are little affected by the use of experimental data for
the pair $^{25}$Mg-$^{25}$Na. The present calculated rates with USDB are
close to the rates given in \citet{Oda}. 

In case of $^{23}$Na, calculated $B(GT)$ values for the transition to the
ground state of $^{23}$Ne are smaller than the experimental value by 
factors 2.67 and 1.96 for USDB and USD, respectively.
Use of experimental $B(GT)$ values enhance the transition rates for the
pair $^{23}$Na-$^{23}$Ne about by a factor of 2.5.
When experimental data are taken into account, the rates for USDB become 
very close to those by \citet{Oda} as shown in Table 1.
Effects of the use of the experimental $B(GT)$ values are important.
Though the rms deviation of calculated GT matrix elements from experimental ones is as small as 0.114 for USDB with the quenching factor \citep{RB},
experimental $B(GT)$ values as well as experimental excitation energies are taken into account here \citep{Tab} when they are available \citep{NND,Till,Endt}. 


\begin{figure*}[bht]
\hspace*{-8mm}
\includegraphics[scale=1.10]{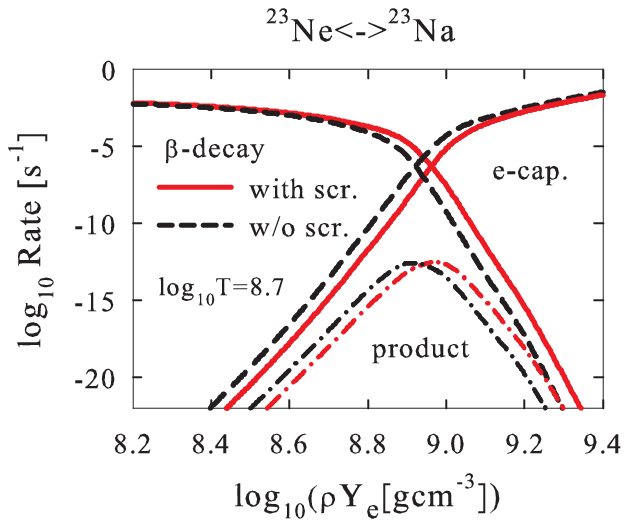}
\hspace*{-13mm}
\includegraphics[scale=1.10]{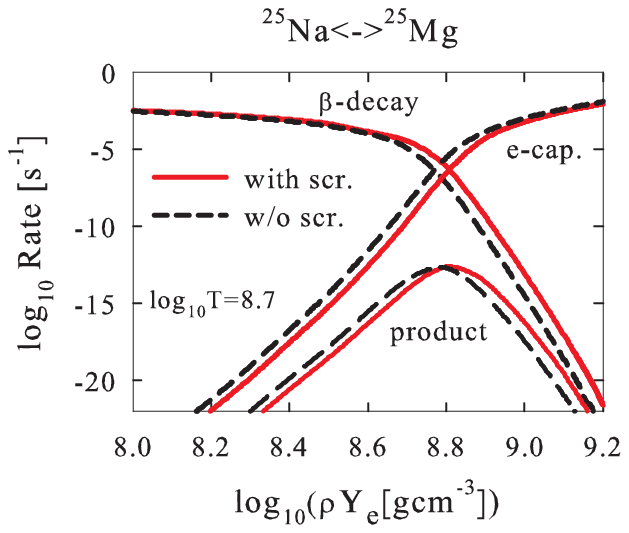}
\caption{
Coulomb screening effects on electron capture and $\beta$-decay rates
for the A=23 URCA nuclear pair, $^{23}$Ne -$^{23}$Na, and A=25 URCA nuclear pair, $^{25}$Na -$^{25}$Mg.
Solid and dashed
curves are obtained with and without the screening effects, respectively,
at log$_{10}T$ =8.7.
The product of the electron capture and the $\beta$-decay rates is also shown.
\label{fig:fig4}}
\end{figure*}

Effects of the Coulomb corrections on the transition rates are studied for the nuclear pairs,
$^{23}$Ne -$^{23}$Na and $^{25}$Na -$^{25}$Mg.
The calculated rates for a temperature log$_{10}T$ =8.7 are shown in Fig. 4 for the pairs
with and without the Coulomb effects.
The URCA density is shifted toward a higher density region by $\Delta$log$_{10}(\rho Y_e)$ =0.04.
Though the change of the
URCA density is rather modest,
it can affect the
fate of the stars with mass around 9M$_{\odot}$.

\begin{figure*}[tbh]
\hspace*{-8mm}
\includegraphics[scale=1.10]{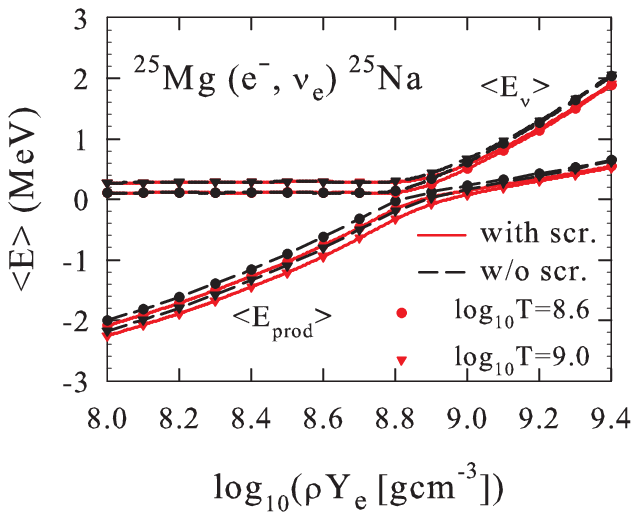}
\hspace*{-13mm}
\includegraphics[scale=1.10]{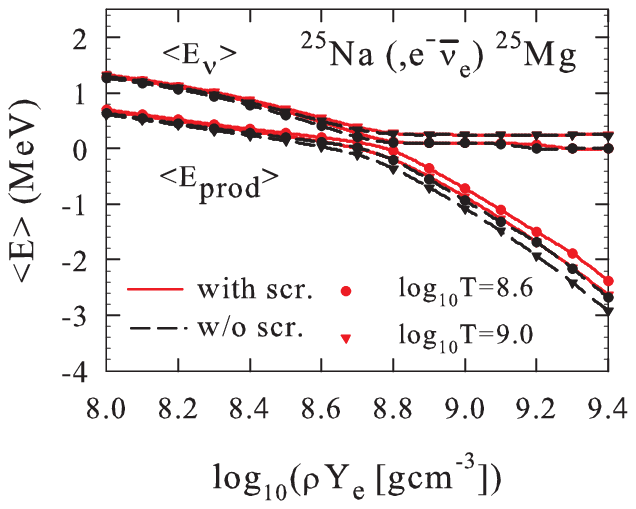}
\caption{
Averaged neutrino energy, $<E_{\nu}>$, and averaged energy production, $<E_{\rm prod}>$, in
electron capture reactions on $^{25}$Mg and $\beta$-decay transitions from $^{25}$Na for
temperatures log$_{10}T$ =8.6 and 9.0  as functions of density
log$_{10}(\rho Y_e)$.
Cases with and without the screening effects are denoted by solid and dashed curves, respectively.
\label{fig:fig5}}
\end{figure*}

\begin{figure*}[tbh]
\hspace*{-8mm}
\includegraphics[scale=1.10]{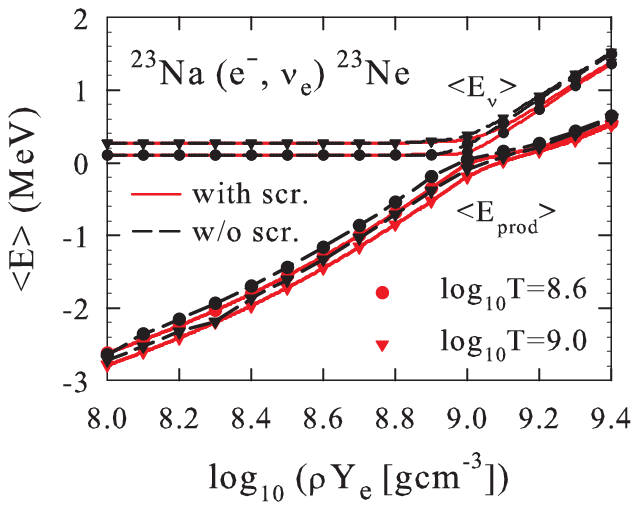}
\hspace*{-13mm}
\includegraphics[scale=1.10]{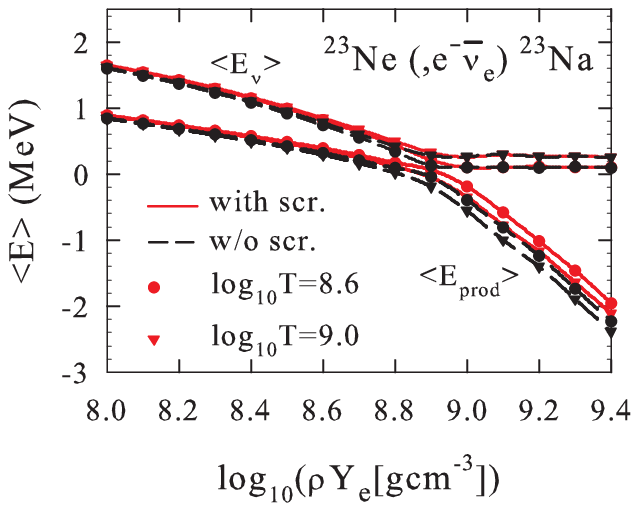}
\caption{
The same as in Fig. 4 for electron capture reactions on $^{23}$Na and $\beta$-decay transitions from $^{23}$Ne.
\label{fig:fig6}}
\end{figure*}

Now we discuss energy loss by neutrino emissions as well as heating by $\gamma$ emissions during the
electron capture and $\beta$-decay processes.
Averaged energy of emitted neutrinos defined by Eq. (7), $< E_{\nu} >$, and averaged energy production
\begin{equation}
< E_{\rm prod} > = \mu_e -Q_{\rm nucl} -<E_{\nu}>
\end{equation}
with $Q_{\rm nucl}$ = $M_{\rm d}c^2$ -$M_{\rm p}c^2$, which is the energy difference between the g.s.'s of daughter
and parent nuclei, for electron capture processes and
\begin{equation}
< E_{\rm prod} > = Q_{\rm nucl} -\mu_e -<E_{\nu}>
\end{equation}
for $\beta$-decay processes are shown in Figs. 5 and 6 for $^{25}$Na -$^{25}$Mg and $^{23}$Ne -$^{23}$Na
pairs, respectively.
The energy generation in the star is determined by \citep{Pinedo14}
\begin{equation}
kT\frac{ds}{dt} = \frac{dY_e}{dt} <E_{\rm prod}>
\end{equation}
where $s$ is the entropy per nucleon.
The condition that the time scale for thermodynamical equilibrium is shorter than the time scale for
the weak interaction is assumed to be satisfied.

Averaged neutrino energies, $<E_{\nu}>$, and averaged energy productions, $<E_{\rm prod}>$, for
electron capture reactions on $^{25}$Mg and $^{23}$Na and $\beta$-decay transitions from $^{25}$Na and $^{23}$Ne
are shown in Figs. 5 and 6.
The value of $<E_{\nu}>$ is almost determined by the values of the electron chemical potential
and the threshold energy $Q_{\rm nucl}$.
In case of electron capture reactions on $^{25}$Mg and $^{23}$Na, the neutrino energy-loss increases above
the URCA densities and the increase of the averaged energy production starts to be suppressed just at the densities,
where the energy production becomes positive.
In case of $\beta$-decay transitions from $^{25}$Na and $^{23}$Ne, the energy production is suppressed due
to neutrino emissions below the URCA densities, while when it becomes negative at the URCA densities the
negative energy production, that is, the energy loss begins to increase monotonically as the density increases.
In both cases of electron capture and $\beta$-decay processes, the energy production is suppressed by neutrino
emissions when it becomes positive.
In case with the Coulomb effects, $Q_{\rm nucl}$ is replaced by $Q_{\rm nucl}$ +$\Delta Q_{\rm C}$ and $\mu_e$ by $\mu_e -V_s(0)$. 
As $\Delta Q_{\rm C}$ is positive and its magnitude is larger than $|V_s(0)|$, $<E_{\rm prod}>$ gets reduced
(enhanced) for the electron capture ($\beta$-decay) case.
This change of $<E_{\rm prod}>$, however, is suppressed above (below) the URCA densities for
the electron capture ($\beta$-decay) case since $<E_{\nu}>$ with the screening effects is smaller
(larger) than that without the effects. 
These featurs are seen in Figs. 5 and 6. 
We will show calculated results with the Coulomb effects hereafter.

\subsection{$^{23}$F -$^{23}$Ne, $^{25}$Ne -$^{25}$Na and $^{27}$Na -$^{27}$Mg pairs}

After the cooling of the O-Ne-Mg core of the stars occur by nuclear URCA processes for pairs with A = 25 and 23,
successive electron capture reactions are expected to be triggered at higher densities corresponding to the Q-values.
Electron capture reactions on $^{24}$Mg with Q = 5.52 MeV is considered to be important in later stages of the star
evolutions leading to electron-capture supernova explosions.


\begin{figure*}[tbh]
\includegraphics[scale=1.5]{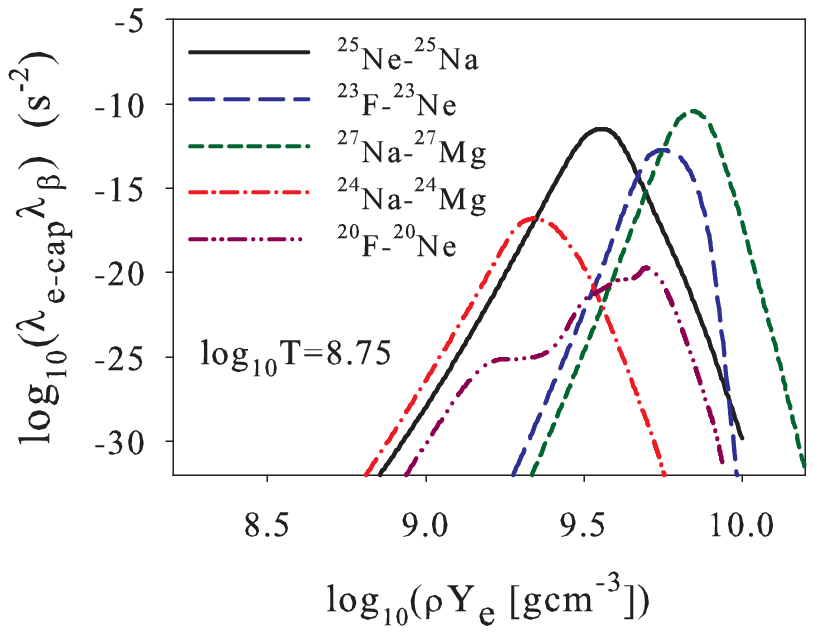}
\caption{
The product of electron capture and $\beta$-decay rates are shown for the nuclear pairs,
$^{25}$Ne -$^{25}$Na, $^{23}$F -$^{23}$Ne, $^{27}$Na -$^{27}$Mg, $^{24}$Na -$^{24}$Mg
and $^{20}$F -$^{20}$Ne for a temperature log$_{10}T$ =8.75 as functions of density
log$_{10}(\rho Y_e)$.
\label{fig:fig7}}
\end{figure*}

The product of electron capture and $\beta$-decay rates for various nuclear pairs are shown in Fig. 7.
The peak position for each pair is determined by the Q-value of each transition.
The URCA densities for the pairs, $^{24}$Na -$^{24}$Mg, $^{25}$Ne -$^{25}$Na,
$^{23}$F -$^{23}$Ne and $^{27}$Na -$^{27}$Mg, increase in this order.
For the pair $^{20}$F -$^{20}$Ne, GT transition can not occur between 
the g.s.'s
as the g.s.
of $^{20}$F is 2$^{+}$ while that of $^{20}$Ne is 0$^{+}$.
In this case, the contributions from the forbidden transitions
betwen the g.s.'s 
are taken into account \citep{Pinedo14}. 
The peak at lower density at log$_{10}(\rho Y_e)\approx$ 9.2 corresponds to the transition
between $^{20}$Ne (2$^{+}$) and $^{20}$F (2$^{+}$) while the peak at higher density at log$_{10}(\rho Y_e)\approx$ 9.7
corresponds to the transition between $^{20}$Ne (0$^{+}$) and $^{20}$F (1$^{+}$).
The peak at log$_{10}(\rho Y_e)\approx$ 9.5 corresponds to the forbiden transition between the ground states.
For $^{24}$Mg and $^{20}$Ne, the heating of stars occur by double electron
capture reactions due to the pairing effects, that is, by successive electron 
capture reactions on $^{24}$Na and $^{20}$F following the capture reactions 
on $^{24}$Mg and $^{20}$Ne, respectively.   


\begin{figure*}[tbh]
\hspace*{-8mm}
\includegraphics[scale=1.10]{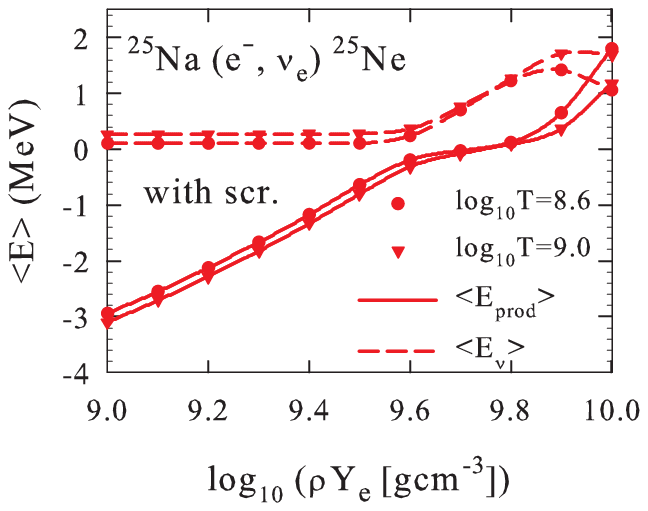}
\hspace*{-13mm}
\includegraphics[scale=1.10]{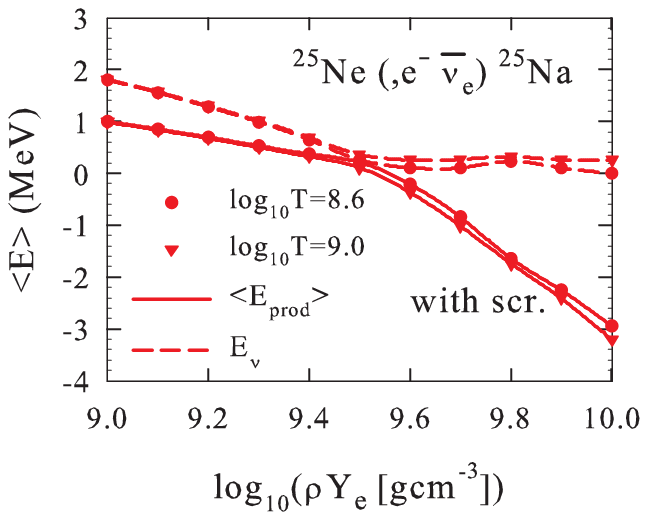}
\caption{
The same as in Fig. 4 for electron capture reactions on $^{25}$Na and $\beta$-decay transitions from $^{25}$Ne.
The Coulmb corrections (screening effects) are included.
Evaluated values for $<E_{\rm prod}>$ and $<E_{\nu}>$ are denoted by solid and dashed curves, respectively.
\label{fig:fig8}}
\end{figure*}

Averaged neutrino energies, $<E_{\nu}>$, and averaged energy productions, $<E_{\rm prod}>$, for
electron capture reactions on $^{25}$Na and $\beta$-decay transitions from $^{25}$Ne
are shown in Fig. 8.
Here, transitions from 5/2$^{+}_{\rm g.s.}$, 3/2$^{+}$ and 1/2$^{+}$ states of $^{25}$Na are included for
the electron capture reactions, while those from 1/2$^{+}_{\rm g.s.}$, 5/2$^{+}$ and 3/2$^{+}$ states of $^{25}$Ne are
included for the $\beta$-decays.

For the $^{25}$Ne -$^{25}$Na, $^{23}$F -$^{23}$Ne and $^{27}$Mg -$^{27}$Na pairs, general features of the neutrino energy-loss and energy production are similar to the cases of
$^{25}$Na -$^{25}$Mg and $^{23}$Ne -$^{23}$Na pairs.
Averaged energy produced is suppressed by neutrino emissions above (below) the URCA densities for electron capture
($\beta$-decay) processes.

When the production yield of $^{24}$Mg from $^{12}$C +$^{12}$C reactions is small and
the contribution of electron capture reactions on $^{24}$Mg to the evolution of stars is suppressed,
the URCA processes for the present pairs can be important.
In a recent model of O-Ne-Mg white dwarfs, these secondary coolings induced by the pairs give
important effects on the evolution of the stars \citep{Sam13,Sam}.


We briefly comment on electron capture and $\beta$-decay rates for nuclear pairs with A=20 and 24,
$^{20}$O -$^{20}$F, $^{20}$F -$^{20}$Ne, and $^{24}$Ne -$^{24}$Na, $^{24}$Na -$^{24}$Mg.
These weak rates are important for late stages of the evolution of stars with 8-10 M$_{\odot}$.
Electron capture reactions on $^{24}$Mg and $^{20}$Ne are important processes toward electron-capture
supernova explosions or Fe core-collapse supernova explosions after contraction
of the electron-degenerate O-Ne-Mg cores. 

The weak rates for these nuclei are examined in detail in \citet{Pinedo14}
by taking into account the forbidden transitions between $^{20}$Ne (0$_{\rm g.s.}^{+}$) and $^{20}$F (2$_{\rm g.s.}^{+}$).
The transitions are found to give non-negligible contributions at log$_{10}T <$ 9.0
in a density region; 9.3$<$ log$_{10}(\rho Y_e) <$9.6. 
Calculated rates with the inclusion of the forbidden transitions are also tabulated in  \citet{Tab}. 
The rates were also examined in \citet{Taka} with the use of USD and available 
experimental data, but the effects of the forbidden transitions were not taken
into account. The GT strength for 0$_{\rm g.s.}^{+}$ $\rightarrow$ 1$^{+}$ (1.057 MeV) obtained from recent experimental (p, n) data on $^{20}$Ne and used in \citet{Pinedo14} were not also taken into account as the data were not available in
1989. 
The GT strength, enhanced in the (p, n) data compared with the USD value, 
enhances the electron capture rates on $^{20}$Ne at log$_{10}(\rho Y_e) >$9.6 about by 60-70$\%$.    

As the Q value for the $^{24}$Ne -$^{24}$Na pair is as small as Q = 2.47 MeV due to the pairing effects,
electron capture reactions on $^{24}$Na occur sucessively after the electron capture process on $^{24}$Mg,
and contribute to
heating the stars at later stages
of evolutions leading to the electron-capture supernova or Fe-core formation. 
The situation is the same as in the successive elecron capture reaction on 
$^{20}$F after the capture reaction on $^{20}$Ne.

\section{Summary}

We have studied electron capture and $\beta$-decay processes relevant to stars with O-Ne-Mg cores.
Nuclear weak rates for $sd$-shell nuclei in stellar environments are evaluated by
shell-model calculations with the use of the USDB Hamiltonian.
The weak rates for nuclear pairs with A=23 and 25, $^{23}$Ne -$^{23}$Na and $^{25}$Na -$^{25}$Mg,
are important for nuclear URCA processes and cooling of the O-Ne-Mg core of the stars.
Evaluations of the rates with fine meshes of density and temperature are required to get
clear URCA densities \citep{TSN}.
Coulomb effects due to electron screening and change of chemical potential of ions with
electron background are evaluated for the weak transitions.
The effects lead to  enhancements of the threshold energy
for electron capture reactions and Q-value for the $\beta$-decays.
The Coulomb corrections shift the URCA densities
toward a higher density region.

Triggering of electron capture reactions is determined by the threshold energy of the
transitions and chemical potential of electrons at high densities and temperatures.
The capture reactions occur in order from pairs with smaller Q-values.
After the first cooling of stars by the URCA processes for the pairs, $^{25}$Na -$^{25}$Mg and $^{23}$Ne -$^{23}$Na,
electron capture reactions on $^{24}$Mg, $^{25}$Na, $^{23}$Ne and $^{27}$Mg ocuur
successively in this order as the density increases and the chemical potential of electron
is increased and reach the corresponding threshold energy.
Secondary cooling of stars by the URCA processes in these nuclei can be important at higher densities \citep{Sam13}.

The neutrino energy-loss rate and $\gamma$-ray heating rate are also evaluated by shell-model calculations.
The averaged neutrino energies and averaged energy production in the electron capture and $\beta$-decay processes
are obtained for the pairs, and the effects of energy loss by neutrino emissions are examined.
Energy production is found to be suppressed by neutrino emissions above (below) the URCA densities for electron capture
($\beta$-decay) processes.


Usually, the heating process by electron capture reactions on $^{24}$Mg and $^{20}$Ne occur after the cooling by
the URCA processes in the $^{25}$Na -$^{25}$Mg and $^{23}$Ne -$^{23}$Na pairs.
If this heating was not enough by some reason, for example, by a lack of enough amount of $^{24}$Mg,
the secondary cooling by URCA processes in $^{25}$Ne -$^{25}$Na,
$^{23}$F -$^{23}$Ne and $^{27}$Na -$^{27}$Mg pairs might have important roles in the evolution of stars \citep{Sam13,Sam}.

Accurate evaluations of the nuclear weak rates at stellar environments are quite important for the studies of
evolution of stars with 8-10 M$_{\odot}$. 
The weak rates determine the final fates of O-Ne-Mg cores and related white dwarfs but also a wider range
of lower mass stars with C-O cores, which contain primordial species of A =20-27. 
For future advances for the studies,
the electron capture and $\beta$-decay rates, neutrino energy-loss rates and $\gamma$-ray heating rates for $sd$-shell
nuclei with A=17-28 have been evaluated with the Coulomb effects and tabulated for densities log$_{10}(\rho Y_e)$ = 8.0 -11.0 and temperatures
log$_{10}T$ = 7.0 -9.65 with fine meshes \citep{Tab}.
The weak-rate data for the pairs with A =23 and 25 have been applied to study the cooling of O-Ne-Mg cores
by the nuclear URCA processes and evolution of stars with 8--10\, M$_{\odot}$ \citep{Jones}.
Stars with $M_{\rm I}=$ 8.8M$_{\odot}$ are found to end up with electron-capture supernovae due to
the nuclear URCA processes, while stars with $M_{\rm I}=$ 9.5M$_{\odot}$ are found to evolve to Fe core-collapse
supernovae.
Evolution of hybrid C-O-Ne white dwarfs as progenitors of type Ia supernovae has been also studied with 
the nuclear URCA processes \citep{Sam}. The hybrid progenitor models might explain the observed diversity of 
Type Ia supernovae. 
 
We notice from Fig. 1 that
the $\beta$-decay Q value for $^{31}$Si $\rightarrow$ $^{31}$P is as small as 1.49 MeV.
The $\beta$-decay Q value for $^{29}$Al $\rightarrow$ $^{29}$Si is also as small as 3.69 MeV.
Electron capture processes on $^{31}$P and $^{29}$Si could have some important roles on the cooling and
neutronization of the core
in later O burning and Si burning stages in more massive stars.

\acknowledgements

The authors would like to thank Sam Jones
for useful communications on evolutions of stars with O-Ne-Mg core.
This work has been supported in part by Grants-in-Aid for Scientific
Research
(S)23224008,
(S)21540267, (C)22540290, (C)26400222, and (C)15k05090,
and the World Premier International Research Center Initiative
of the MEXT of Japan.

\end{document}